\documentclass{PoS}

\usepackage{graphicx,times}
\usepackage{epsf}
\usepackage{amsmath,amssymb,latexsym,float}
\usepackage{enumerate}
\usepackage{listings}
\numberwithin{equation}{section}

\newcommand{\R}{\text{\fontshape{n}\selectfont I\kern-.42exR}}

\newcommand{\1}{\text{\fontshape{n}\selectfont 1\kern-.56exl}}

\title{Minimally Doubled Fermion Revival}

\ShortTitle{Minimally Doubled Fermion Revival}

\author{\speaker{Artan Bori\c{c}i}\\%\thanks{A footnote may follow.}\\
        Faculty of Natural Sciences\\
        University of Tirana\\
        King Zog I Boulevard, Tirana, Albania\\
        \email{borici@fshn.edu.al}}

\abstract{
\vspace{1cm}
In this paper, we present the recent progress on minimally doubled lattice actions. In particular, we discuss the proposal of Creutz and its variations on an orthogonal lattice. A preliminary computation of the pion mass on an SU(3) background field shows the expected behaviour as predicted form the chiral perturbation theory.
}

\FullConference{The XXVI International Symposium on Lattice Field Theory\\
		 July 14-19 2008\\
		 Williamsburg, Virginia, USA}

\begin{document}

\section{Introduction}

Motivated by the Dirac structure of the graphene electrons in two dimensions, Creutz introduced a lattice action, which in four dimensions, describes two flavours of Dirac fermions with exact chiral symmetry \cite{Creutz_07}. In momentum space, this action has the form:
\begin{equation}\label{Creutz_operator}
D(p)=i~B\gamma_4(4C-\sum_{\mu}\cos p_{\mu}) + i~\sum_{k=1}^3\gamma_k s_k(p)\ ,
\end{equation}
where
$$
\begin{matrix}
s_1(p)=\sin p_1+\sin p_2-\sin p_3-\sin p_4\\
s_2(p)=\sin p_1-\sin p_2-\sin p_3+\sin p_4\\
s_3(p)=\sin p_1-\sin p_2+\sin p_3-\sin p_4
\end{matrix}
$$
and $B,C$ are free parameters. Creutz has shown that the zeros of this operator in the Brillouin zone are at $(\tilde p,\tilde p,\tilde p,\tilde p)$ and $(-\tilde p,-\tilde p,-\tilde p,-\tilde p)$, where $C=\cos \tilde p$. In order to get a lattice action with one zero at the origin, we proposed a formulation, which shifts the zeros in the Brillouin zone at $(0,0,0,0)$ and $(\tilde 2p,\tilde 2p,\tilde 2p,\tilde 2p)$ \cite{Borici_08}.

The minimal number of zeros allowed by the Nielsen-Ninomiya theorem is exactly two \cite{NN_1981}. Hence, these actions achieve a minimal doubling of fermion species on the lattice. In fact, the idea of minimally doubled action is not new. It was first pointed out by Karsten and later by Wilczek \cite{Karsten_1981,Wilczek_1987}. Their actions are unitary equivalent to each other. In the free case, the Wilczek action has the form:
$$
D(p)=\sum_{\mu=1}^4~i\gamma_{\mu}~\sin p_{\mu} +i\gamma_4\sum_{k=1}^3(1-\cos p_k)\ .
$$
It has two zeros: one at $(0,0,0,0)$ and the other at $(0,0,0,\pi)$.

\section{The Creutz Action on Orthogonal Axes}

The easiest way to get a lattice action with one zero at the origin is to put the Creutz action on an orthogonal lattice, where the action parameters satisfy $BS=C$, where $S=\sin \tilde p$. Setting $B=1$ and translating momenta, $p_{\mu}=\tilde p+q_{\mu}$, we get:
$$
D(q)=i\gamma_4(4C-C\sum_{\mu}\cos q_{\mu}) + iC~\sum_{k=1}^3\gamma_k s_k(q)+iS\gamma_4\sum_{\mu}\sin q_{\mu}+iS~\sum_{k=1}^3\gamma_k c_k(q)\ ,
$$
where we have denoted:
$$
\begin{matrix}
c_1(q)=(\cos q_1-1)+(\cos q_2-1)-(\cos q_3-1)-(\cos q_4-1)\\
c_2(q)=(\cos q_1-1)-(\cos q_2-1)-(\cos q_3-1)+(\cos q_4-1)\\
c_3(q)=(\cos q_1-1)-(\cos q_2-1)+(\cos q_3-1)-(\cos q_4-1)
\end{matrix}\
$$
Taking $C=1/\sqrt{2}$ as well as $S+C=0$, and defining:
\begin{eqnarray*}
s_4(q)&=&-\sin q_1-\sin q_2-\sin q_3-\sin q_4\\
c_4(q)&=&(\cos q_1-1)+(\cos q_2-1)+(\cos q_3-1)+(\cos q_4-1)\ ,
\end{eqnarray*}
the {\it p-translated} Creutz action takes the form:
$$
D(q)=\sum_{\mu}i\gamma_{\mu} s_{\mu}(q) + \sum_{\mu}i\gamma_{\mu} c_{\mu}(q)\ ,
$$
or in the scalar product notation, $(\gamma,x)=\sum_{\mu}\gamma_{\mu}x_{\mu}$, one has:
$$
D(q)=i(\gamma,s(q)+c(q))\ .
$$
Using the following orthogonal matrices:
$$
a:=\frac12
\begin{pmatrix}
 1 &  1 & -1 & -1 \\
 1 & -1 & -1 &  1 \\
 1 & -1 &  1 & -1 \\
-1 & -1 & -1 & -1 \\
\end{pmatrix}\ ,
~~~~~~b:=-\frac12
\begin{pmatrix}
 1 &  1 & -1 & -1 \\
 1 & -1 & -1 &  1 \\
 1 & -1 &  1 & -1 \\
 1 &  1 &  1 &  1 \\
\end{pmatrix}\ ,
$$
and noting that,
$$
s=2a\tilde s, ~~~~~~c=2b\tilde c\ ,
$$
where
$$
\tilde s=(~\sin q_1,~\sin q_2,~\sin q_3,~\sin q_4~)^T, ~~~~\tilde c=(\cos q_1 - 1,~\cos q_2 - 1,~\cos q_3 - 1,~\cos q_4 - 1~)^T\ ,
$$
then, the rescaled action by a factor of $2$ can be written in the form:
$$
D(q):=i(\gamma,a\tilde s(q)+b\tilde c(q))=i(a^T\gamma,\tilde s(q)+a^Tb\tilde c(q))\ .
$$
Denoting,
$$
\alpha:=a^Tb=\frac12
\begin{pmatrix}
-1 &  1 &  1 &  1 \\
 1 & -1 &  1 &  1 \\
 1 &  1 & -1 &  1 \\
 1 &  1 &  1 & -1 \\
\end{pmatrix}\ ,
$$
we get:
$$
D(q)=i(a^T\gamma,\tilde s(p)+\alpha\tilde c(q))\ .
$$
It is easy to show that $a^T\gamma$ are Dirac gamma matrices:$P_L$ and $P_R$
$$
\{(a^T\gamma)_{\mu},(a^T\gamma)_{\nu}\}=\sum_{\rho,\sigma}a_{\rho\mu}a_{\sigma\nu}\{\gamma_{\rho},\gamma_{\sigma}\}=2\sum_{\rho}a_{\rho\mu}a_{\rho\nu}=2\delta_{\mu\nu}\ .
$$
Therefore, the factor $a^T$ can be dropped. This way, the final expression has the form:
\begin{eqnarray*}
D(q)&=&i(\gamma,\tilde s(q))+i(\gamma',\tilde c(q))\\
&=&\sum_{\mu}~i\gamma_{\mu}~\sin~q_{\mu}+\sum_{\mu}~i\gamma'_{\mu}(\cos~q_{\mu}-1)\ ,
\end{eqnarray*}
where $\gamma'=\alpha\gamma$ are again Dirac gamma matrices for the same reason as above. Noting that,
$$
\sum_{\mu}\gamma_{\mu}=\sum_{\mu}\gamma'_{\mu}\equiv2\Gamma\ ,
$$
we get another expression for the fermion action:
\begin{equation}\label{final_form}
D(p)=\sum_{\mu}~i\gamma_{\mu}~\sin~q_{\mu}+\sum_{\mu}~i\gamma'_{\mu}\cos~q_{\mu}-2i\Gamma\ .
\end{equation}
This expression was elegantly derived by Creutz in terms of a linear combination of two naive actions plus the $-2i\Gamma$ term, the latter cancelling exactly one naive action at the zeros of $D$ \cite{Creutz_08}.

\subsection{The Dirac Operator in Position Space}

In order to write down the Dirac operator in position space, we express the momentum space operator in terms of forward and backward propagating plane waves, $e^{ip_{\mu}}$ and $e^{-ip_{\mu}}$:
$$
D(p)=m+\frac{i}{2}\sum_{\mu}\left[\left(\gamma'_{\mu}-i\gamma_{\mu}\right)e^{ip_{\mu}}+\left(\gamma'_{\mu}+i\gamma_{\mu}\right)e^{-ip_{\mu}}\right]-2i\Gamma\ ,
$$
where we have added the bare fermion mass, $m$. Then, by making the formal substitution $ip_{\mu}\rightarrow \partial_{\mu}$, one gets:
$$
D=m\1+\frac{i}{2}\sum_{\mu}\left[\left(\gamma'_{\mu}-i\gamma_{\mu}\right)e^{\partial_{\mu}}+\left(\gamma'_{\mu}+i\gamma_{\mu}\right)e^{-\partial_{\mu}}\right]-2i\Gamma\ ,
$$
where the shift operators, $e^{\partial_{\mu}}$ and $e^{-\partial_{\mu}}$, are defined by their action on a Dirac field $\psi(x)$:
$$
e^{\pm\partial_{\mu}}\psi(x)=\psi(x\pm a\hat\mu)\ .
$$
Hence, the position space Dirac operator can be implemented using the following terms:
\begin{itemize}
\item the on-site term, $(m\1-2i\Gamma)\psi_i$.
\item the forward hopping term, $\frac{i}{2}(\gamma'_{\mu}-i\gamma_{\mu})\psi(x+a\hat\mu)$.
\item the backward hopping term, $\frac{i}{2}(\gamma'_{\mu}+i\gamma_{\mu})\psi(x-a\hat\mu)$.
\end{itemize}
As usual, the gauge fields are introduced by requiring the hopping terms to be gauge covariant.

\section{Minimally Doubled Actions and Hypercubic Symmetry}

Shortly after our proposal, it was noted that the lattice action given above lacks the full symmetry of the hypercubic group \cite{Bedaque_et_al}. The reason is that the action picks as a special direction the main diagonal of the hypercube. Hence, in the presence of gauge field interaction, there is a dimension five operator that enters the action, namely $\bar\psi\Gamma\nabla^2\psi$ \cite{Creutz_08}.

As we pointed out earlier, the idea of minimally doubled action is not new. The actions of Karsten and Wilczek pick the time axis as a special direction. Again, the loss of the full hypercubic symmetry introduces extra relevant terms in the interacting case \cite{Wilczek_1987,Bedaque_et_al}.

Since a minimally doubled action has necessarily two zeros, the line that joins the two zeros in the Brillouin zone defines a special direction. Hence, independently of a specific action, the hypercubic symmetry will be broken. In principle, there is nothing special about a hypercubic action. The authors of reference \cite{Bedaque_et_al2} propose a `hyperdiamond' action, which generalises the graphene structure in five dimensions. However, the resulting action has more than two zeros. It is an interesting question whether it is possible to have a minimally doubled action which does not break the original symmetry of the action.

\section{A Preliminary Numerical Test}

\vspace{-0.1cm}
So far, the present formulation has only been tested at tree level perturbation theory \cite{Cichy_et_al}. It has been shown that the scaling violations are of the order $O(a^2)$, as expected. Here, we present preliminary results on the pion mass calculations on ten $16^332$ lattices generated with the SU(3) Wilson gauge action at $\beta=6$.

\subsection{The Quark Propagator}

\vspace{-0.15cm}
The quark propagator, $g$, is computed as the solution of linear system $Dg=\delta$, where $\delta$ is taken to be a point source. Since the massless operator is antihermitian, it is easy to see that $D$ is normal, i.e. $D^*D=DD^*$. This property, which is shared by the Kogut-Susskind operator \cite{KS_1975}, allows one to use optimal inversion algorithms. Indeed, the Conjugate Gradients algorithm on Normal Equations (CGNE) and the Conjugate Residual (CR) algorithm are optimal Krylov subspace based algorithms for staggered fermions \cite{Borici_PhDthesis}. Hence, the minimally doubled fermion presented here shares the same numerical advantages, while describing two species of fermions instead of four. For the quark propagator computations we have used the CGNE algorithm.

\vspace{.1cm}
\hspace{1cm}\includegraphics[scale=0.6]{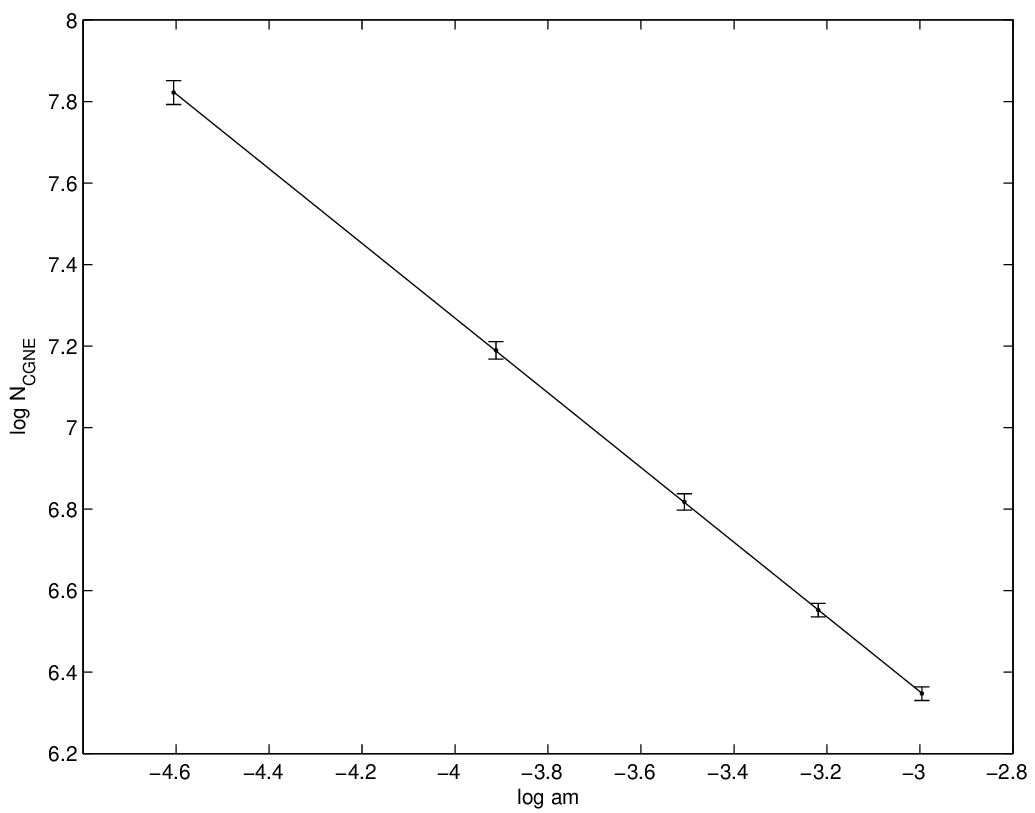}

{\bf Figure 1}. The scaling law of CGNE iterations with the bare quark mass. The critical exponent is computed to be $z=0.92(4)$.

\vspace{.3cm}

While optimal, this algorithm suffers from the critical slowing down, as all Krylov subspace-based inversion algorithms do. Thus, the the number of CGNE iterations to reach a fixed accuracy is expected to scale like $\sim 1/(am)^z$ with the inverse quark mass, $am$. Our data suggest a critical exponent value $z=0.92(4)$, which is clearly  smaller than 1, the expected value for Wilson fermions \cite{Borici_PhDthesis}.

In the present calculations, we haven't made any effort to accelerate the inversion. The even-odd acceleration does not work for these type of fermions since the on-site term does not commute with the hopping term.

\subsection{The Pion Propagator and the Pion Mass}

\vspace{-0.15cm}
Zero momentum pion propagators are computed using the vacuum expectation values of correlation functions of pion interpolating fields, $\bar\psi(x)\gamma_5\psi(x)$:
\begin{eqnarray*}
G(t)&=&\sum_{\vec{x}} <0|\bar\psi(x)\gamma_5\psi(x)\bar\psi(0)\gamma_5\psi(0)|0>\\
&=&\sum_{\vec{x}} g(x,0)^*g(x,0)\ .
\end{eqnarray*}
For large $t$ the right hand side can be fitted to the ground sate ansatz:
$$
G(t) \sim \cosh \text{am}_{\pi}(T/2+1 - t)\ ,
$$
where periodic boundary conditions are applied and $T$ is the lattice extension along the fourth diection. We computed effective masses by inverting the expression:
$$
\frac{G(t+1)}{G(t)}=\frac{\cosh\text{am}_{\pi} (T/2-t)}{\cosh\text{am}_{\pi}(T/2+1 -t)}\ ,
$$
where the the symmetry with respect to the lattice mid point $T/2+1$ is enforced. In figure 2 we plot the pion effective mass squared at different time slices for quark masses $am=0.01,0.02,0.03,0.04,0.05$. Since effective masses show a flat behaviour one can pick the value at a given time slice. We have selected the values at the last time slice which display the largest errors.

\vspace{.1cm}
\hspace{1cm}\includegraphics[scale=0.6]{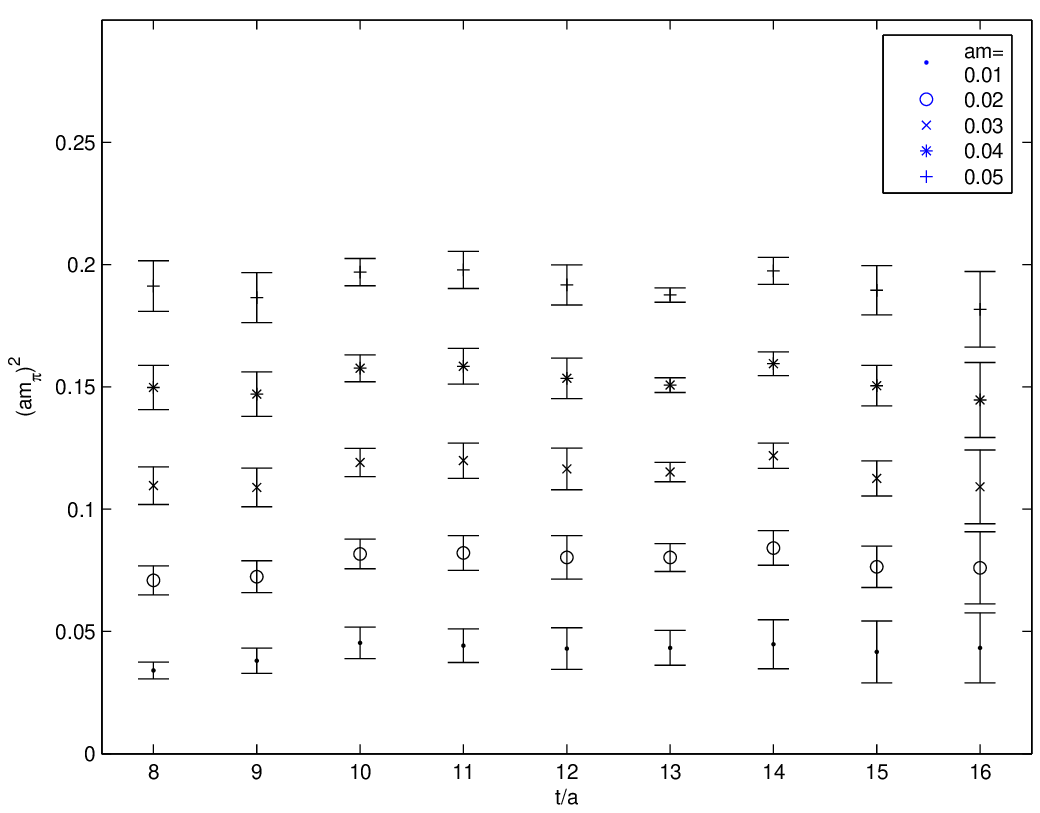}

\hspace{2cm}{\bf Figure 2}. Pion effective masses squared for different quark masses.

\vspace{.3cm}
In figure 3 we show the behaviour of the pion mass squared against the quark mass. The full line is the least squares fit of the data which gives the result:
$$
(am_{\pi})^2=0.007(17)+3.5(4)(am)\ .
$$
The figure shows the extrapolated pion mass at zero quark mass, $0.007(17)$, which in dimensionful units gives a pion mass $m_{\pi}\simeq 170(200)$ MeV. Here, we have assumed that, at this coupling, the inverse lattice spacing is $\sim 2$ GeV. At this accuracy, the pion mass is consistent to zero.

To conclude, we have reviewed the recent efforts to revive the minimally doubled actions. We have made a preliminary calculation of the pion mass on a SU(3) background, which behaves as predicted by the chiral perturbation theory within the statistical error bars. The final results of this ongoing calculation will reveal any possible discrepancy to the chiral perturbation theory. One notices, however, that the invested computational effort to obtain these results is much smaller than for the Ginsparg-Wilson fermions. Therefore, we conclude that the minimally doubled action presented here is worth exploring in the future.

\vspace{.5cm}
\hspace{1cm}\includegraphics[scale=0.55]{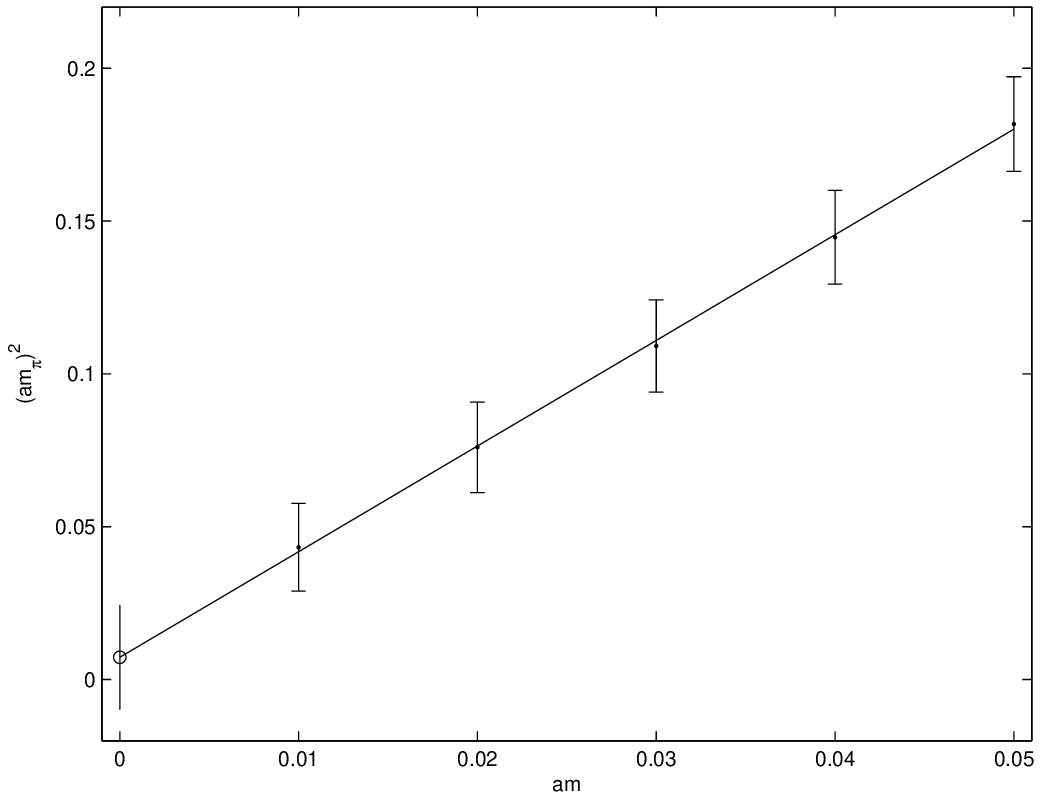}
%\vspace{1cm}

{\small {\bf Figure 3}. Testing chiral perturbation theory with minimally doubled actions: pion mass squared against the bare quark mass. At zero quark mass, the extrapolated pion mass is $m_{\pi}\simeq 170(200)$.}

\end{document}